# Excited-State Trions in a Quantum Well

Sourabh Jain[1], Mikhail Glazov[2], and Ashish Arora[1,*]

[1]Department of Physics, Indian Institute of Science Education and Research, Dr. Homi Bhabha Road, 411008 Pune, India
[2]Ioffe Institute, 26 Polytechnicheskaya, 194021 St. Petersburg, Russia
*Email: ashish.arora@iiserpune.ac.in

We report on the observation of an excited 2s state of a trion in a 4.2 nm wide doped GaAs/Al$_{0.3}$Ga$_{0.7}$As quantum well (QW) using magneto-optical Kerr effect (MOKE) spectroscopy under out-of-plane magnetic fields up to 6 T. This resonance appears slightly below the 2s exciton in energy. Strikingly, the 2s trion is found to be bound only for magnetic fields larger than 1 T. The signature of the 2s trion is absent in the magneto-reflectance spectra, while it is detectable in the MOKE spectra signifying the importance of the powerful technique. Similar to the 1s states, the 2s trion shows an opposite degree of magnetic-field-induced polarization compared to its exciton counterpart, in agreement with our theoretical calculations. This transfer of oscillator strength between the complexes establishes an optical fingerprint of the 2s excited trion.

While the existence of a doubly-excited 2s state of the H$^-$ ion (see Fig. 1) has been well established both theoretically and experimentally [1–3], its counterpart in the solids i.e. the excited-state 2s trion has only been recently discovered in a monolayer of a van der Waals semiconductor WS$_2$ [4]. Following the discovery, there is a significant interest in the experimental as well as theoretical condensed matter physics community where such states have been uncovered in other layered two-dimensional (2D) systems as well such as in monolayers of WSe$_2$ [5,6], MoSe$_2$ [6,7], and MoTe$_2$ [8]. However, existence of excited-state trions in the optical spectra of the most well-studied 'conventional' 2D systems, i.e. group III-V or II-VI quantum wells is a longstanding question since many decades from both theoretical and experimental perspectives [9–16]. This is possibly due to the following reasons: Firstly, it is not known if the excited-state trions in the quantum wells are bound states [10,13]. Even if they are bound, their binding energies are expected to be extremely small given that the corresponding numbers for the ground-state 1s trions are already quite small (1 meV to a few meV) [14,15]. Secondly, oscillator strength (if any) of the excited-state trions is expected to be feeble, since their excited- exciton counterparts have almost negligible oscillator strengths [17,18].

Here, we overcome these challenges and discover an excited state of a trion in a 4.2 nm wide GaAs quantum well by performing highly sensitive magneto-optical Kerr effect (MOKE) spectroscopy. In agreement with our theoretical calculations, the magnetic-field-induced quantum confinement of the excited 2s trions enhances their oscillator strengths as well as binding energies [19]. This makes their detection possible in this work.

MOKE spectroscopy measurements are performed on a 4.2 nm wide GaAs single quantum well with Al$_{0.3}$Ga$_{0.7}$As barriers, grown using metal-oxide vapor phase epitaxy on GaAs (001) substrate. The experimental setup is based on polarization modulation spectroscopy [20,21] and is described in the supplemental material [22]. Kerr rotation $\phi_k$ and ellipticity $\eta_k$ are calculated using the following relationships [20,23]

$$\frac{I_f}{I_{DC}} = 2\sqrt{2}J_1(\delta_0)\eta_k; \quad \frac{I_{2f}}{I_{DC}} = 2\sqrt{2}J_2(\delta_0)\phi_k \qquad (1)$$

where $I_{DC}$, $I_f$ and $I_{2f}$ are the components of the signal measured at 0 frequency (picked up by a DC voltmeter), $f = 42$ kHz and $2f = 84$ kHz measured using two lock-in amplifiers. A factor of $1/\sqrt{2}$ is introduced to the right hand side of Eqs. (1) because a lock-in amplifier measures the RMS value of the signals.

In Fig. 2(a) to (c), gray spheres (solid lines) are the experimental (modeled) magneto-reflectance, Kerr rotation and Kerr ellipticity spectra, respectively, measured under magnetic fields from $B = 1$ T to 6 T. For a quantitative analysis, we model our spectra using a transfer-matrix-based method to account for interferences taking place at the multiple interfaces of the sample [23,24]. We begin by assuming the form of a

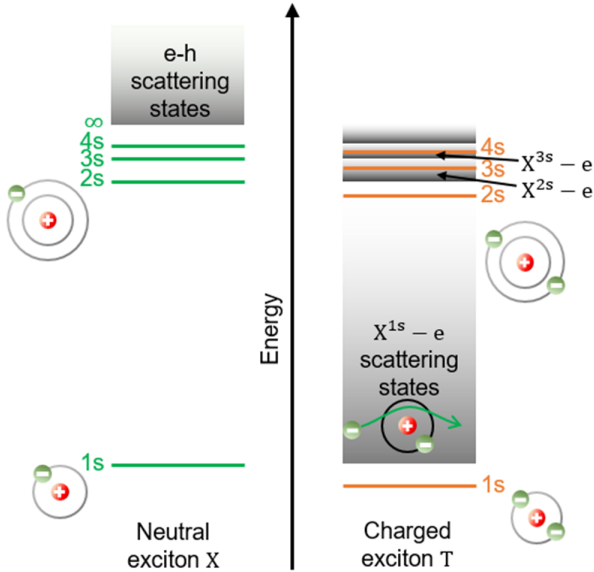

Figure 1. Rydberg series of a neutral exciton X (left side) and a charged exciton (or a trion) T (right side). For neutral exciton X, relative energy positions of the $n = 1$ ground state, $n > 2$ excited states are marked as solid green lines, and e-h scattering states (beyond $n = \infty$) are marked as a shaded gray region. For the trion T, only doubly-excited bound states are shown (orange solid lines) since singly-excited states are unstable. Shaded gray regions depict X − e scattering states.

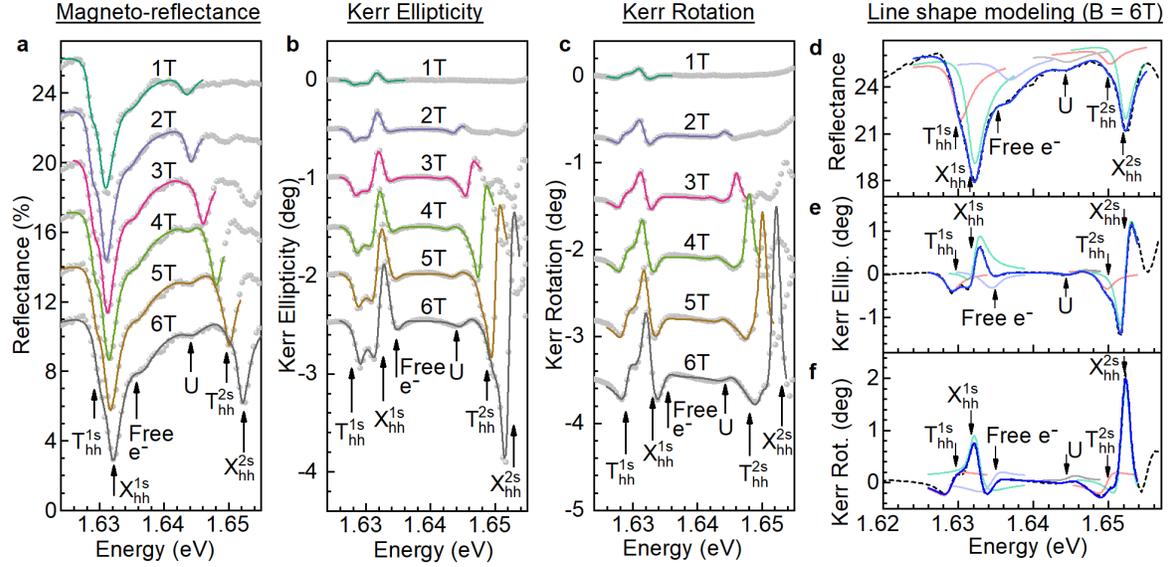

Figure 2. (a) Magneto reflectance, (b) Kerr ellipticity and (c) Kerr rotation spectra of a 4.2 nm wide GaAs single quantum well with $Al_{0.3}Ga_{0.7}As$ barriers, under magnetic fields from $B = 1$ T to 6 T at low temperature of $T = 11$ K. Gray spheres are experimental data and colored solid lines are the modeled spectra. Spectra corresponding to magnetic fields of 2 T to 6 T are shifted vertically below with respect to the 1 T spectra, for clarity. The spectral shifts for each successive spectrum with respect to 1 T spectra in (a) to (c) are of $-3\%$, $-0.5°$ and $-0.7°$, respectively. While 1s states show a small diamagnetic shift, 2s neutral exciton state $X_{hh}^{2s}$ clearly gains oscillator strength and blue shifts strongly with magnetic field as expected. Additionally, a 2s trion state $T_{hh}^{2s}$ emerges as a weak shoulder on the low-energy side of $X_{hh}^{2s}$ in the MOKE spectra, as magnetic field rises. (d), (e) and (f) depict an example of the line shape modeling of the reflectance, Kerr rotation and ellipticity spectra, respectively, for $B = 6$ T. Dashed and solid dark blue lines are the experimental spectra and the modeled spectra, respectively. Light red, green and blue lines represent the individual trion, exciton and free electron contributions of various resonances, respectively. The state U (unidentified origin, contribution represented by light gray solid line) which starts to appear under $B \geq 3$ T magnetic field is not considered for analysis in this work.

dielectric function of the 4.2 nm wide active GaAs layer as follows:

$$\epsilon(E) = (n_b + ik_b)^2 + \sum_j \frac{A_j}{E_j^2 - E^2 - i\gamma_j E} \quad (2)$$

Here, $n_b + ik_b$ is the background refractive index of GaAs, the j$^{th}$ quasiparticle resonance is represented by a complex Lorentzian of energy $E_j$, linewidth $\gamma_j$, and oscillator strength parameter $A_j$. First, we fit the reflectance spectrum such as in Fig. 2d, which requires 6 resonant contributions to represent the observed spectral features. An identification of these resonances is described later based on the fitting parameters obtained. From the modelled reflectance spectrum, we obtain the complex Fresnel reflectance coefficient $\hat{r}(E) = re^{i\theta}$, where $r$ and $\theta$ are functions of energy, $E$. Using this, we simultaneously fit the Kerr rotation and ellipticity spectra [20], which we briefly describe here.

The Kerr rotation $\phi_k$ and ellipticity $\eta_k$ are related to the complex reflectance amplitudes $\hat{r}_\pm(E) = r_\pm e^{i\theta_\pm}$ of the $\sigma^\pm$ circular polarizations as

$$\phi_k(E) = -\frac{1}{2}(\theta_+ - \theta_-), \quad \eta_k(E) = \frac{1}{2}\left(\frac{r_+^2 - r_-^2}{r_+^2 + r_-^2}\right) \quad (3)$$

respectively. In the absence of a magnetic field ($B = 0$), the complex reflectance amplitude obtained from fitting the reflectance spectrum is assumed to be identical for both $\sigma^\pm$ polarizations i.e., $\hat{r}_+ = \hat{r}_-$. In the presence of a magnetic field, the $\sigma^\pm$ polarized components of the j$^{th}$ quasiparticle resonance undergoes a Zeeman splitting $\Delta E_j = E_j^+ - E_j^- = g_j \mu_B B$ ($g_j$ is the effective g factor of the quasiparticle and $\mu_B$ is the Bohr's magneton). Additionally, $\sigma^\pm$ polarized components can also have a different oscillator strength parameters ($\Delta A_j = A_j^+ - A_j^-$) and different linewidths ($\Delta \gamma_j = \gamma_j^+ - \gamma_j^-$) [25,26]. Therefore, $\hat{r}_\pm$ differ leading to non-zero signals in the MOKE spectra [22,23]. Simulations reveal that each of the $\Delta E$, $\Delta A$ and $\Delta \gamma$ parameters result in a specific spectral line shape of the MOKE spectra [22]. The magnitude of the MOKE spectral line increases proportionately with the values of these parameters. We find that in the present work (Fig. 2), only $\Delta E$ and $\Delta A$ are required to model the spectral line shapes satisfactorily.

For the lowest magnetic fields ($B = 1$ T in Figs. 2a – 2c) the spectra are relatively simple to understand, and we discuss these first. In Fig. 2a (reflectance), spectral resonances corresponding to the ground state neutral ($X_{hh}^{1s}$) and singlet charged ($T_{hh}^{1s}$) heavy-hole excitons are observed at energies $E_X^{1s} = 1630.8 \pm 0.1$ meV and $E_T^{1s} = 1628.8 \pm 0.1$ meV, respectively. This corresponds to a trion binding energy of $E_X^{1s} - E_T^{1s} = 2.0 \pm 0.2$ meV, which is in agreement with the literature [15,27]. We notice corresponding features in Kerr ellipticity (Fig. 2b) and rotation (Fig. 2c) as well for $B = 1$ T. In Fig. 2a (reflectance), towards the higher energy side of the $X_{hh}^{1s}$, a feature corresponding to the exciton-electron scattering states $(X_{hh}^{1s} - e^-)$ [28,29] is observed at $E_{X-e} = 1635.0 \pm 0.2$ meV. Further onwards, n = 2 excited state of neutral exciton $X_{hh}^{2s}$ appears as a resonance around $E_X^{2s} = 1643.4 \pm 0.2$ meV. In the MOKE spectra, these resonances start to appear more clearly for $B \geq 2T$ as we discuss in the next para.

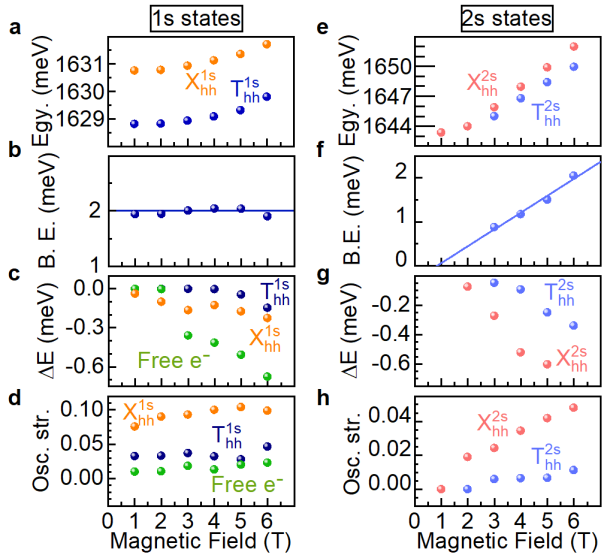

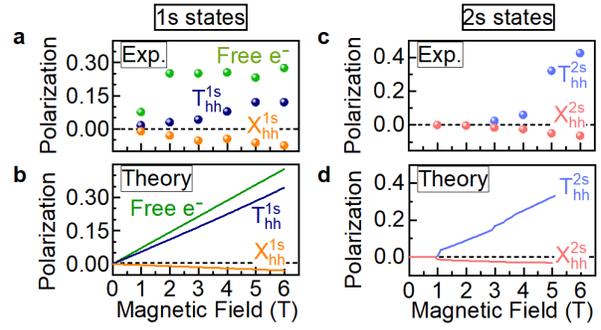

Figure 3. (a) to (d) are the transition energies, trion binding energy, Zeeman splitting, and relative transition oscillator strength, respectively, of the 1s neutral and charged exciton states. (e) to (h) represent the corresponding parameters of the 2s exciton and trions. The parameters are derived from the line shape modeling as described in Figure 2, and the main text. The most notable deduction is the following: In (f), the binding energy of the 2s trion becomes 0 around $B \approx 1$ T when extrapolated on the magnetic field axis, pointing towards an unbound 2s trion state in the absence of a magnetic field.

For $B > 1$ T, 1s and 2s exciton and trion states experience a diamagnetic shift with rising magnetic field, where $X_{hh}^{2s}$ shows the largest shift in agreement with previous works [30–33]. The (excited) 2s trion $T_{hh}^{2s}$ appears only in the MOKE spectra as a low-energy shoulder to the $X_{hh}^{2s}$. The evidence behind its identification becomes clearer as the discussion progresses. Transition energies for 1s and 2s states are plotted in Fig. 3a and 3e showing the expected trends of diamagnetic shifts [30–33]. Notably, the binding energy of the 1s trion $T_{hh}^{1s}$ (i.e. its energetic separation from its exciton counterpart $X_{hh}^{1s}$) remains nearly constant as seen in Fig. 3b. It is expected since the 1s trion binding energy exceeds the cyclotron gap $\hbar\omega_c$ because of which, the magnetic field does not significantly affect the trion internal structure. On the other hand, binding energy of the 2s trion $T_{hh}^{2s}$ in Fig. 3f increases monotonically with increasing magnetic field. A simplified linear extrapolation of the $T_{hh}^{2s}$ binding energy crosses the magnetic field axis around $B = 1$ T. This suggests that the $T_{hh}^{2s}$ state is most likely unbound in the absence of the magnetic field. This is possibly the reason that we are unable to detect $T_{hh}^{2s}$ for magnetic fields up to $B = 2$ T. This observation is in line with the generic model of Ref. [5] where the excited state trion binding energy is given by

Figure 4. (a) and (c) show degrees of polarization $p$, Eq. (5), of 1s and 2s states respectively, derived from line shape modeling illustrated in Fig. 2. (b) and (d) are the calculated degrees of polarizations for 1s and 2s states respectively as described in the main text. The doping concentration used in the calculations is $n = 0.5 \times 10^{10}$ cm$^{-2}$ which reproduces the experimental results quite well.

$$E_{T_{hh}^{2s}}^b = E_{X_{hh}^{2s}}^b \exp\left(\frac{1}{DV_{22}}\right) \cos\phi \quad (4)$$

Here, $E_{X_{hh}^{2s}}^b$ is the 2s exciton binding energy, $D$ is the electron-exciton reduced density of states, $V_{22}$ is the effective matrix element of the 2s exciton – electron attraction. The non-zero phase $\phi = \pi|V_{12}|^2/V_{22}^2$ is related to the electron-induced 2s-1s exciton interaction $V_{12}$, and the coupling with other states (2p, 3s, …) is neglected. In the vicinity of the $T_{hh}^{2s}$ state formation threshold, the magnetic field dependece of the 2s trion binding energy is mainly controlled by the $\phi(B) \approx \frac{\pi}{2} - \beta(B - B_{cr})$ dependence, where $B_{cr}$ is the critical magnetic field at which the $T_{hh}^{2s}$ becomes bound and $\beta > 0$ is the parameter describing the magnetic field dependence of the phase. In this model, the magnetic field dependence of the 2s trion binding energy is linear in the vicinity of $B \approx B_{cr}$.

All of the features gain amplitude in the MOKE spectra (Figs. 2b and 2c). This rise in the Kerr signal is because,

(i) Magnitude of the Zeeman splitting $|\Delta E|$ increases with magnetic field [34,35] (see Figs. 3c and 3g). Zeeman splittings of all the states (n = 1,2 in Figs. 3c and 3g) are found to have the same sign (negative). This hints towards a similar origin of all these states and supports our assignment of the $T_{hh}^{2s}$.

(ii) Resonances are expected to gain oscillator strength under magnetic field [30]. We notice that for $X_{hh}^{1s}$, $X_{hh}^{2s}$ and $T_{hh}^{2s}$, a clear enhancement of the oscillator strength is observed (see Figs. 3d and 3h) which is well documented in literature for the Coulomb-bound resonances.

(iii) Optical weights of $\sigma^\pm$ components $(A^\pm)$ of the resonances differ, and $|\Delta A| = |A^+ - A^-|$ increases with magnetic field [9,36,37]. We define a polarization parameter for the $j^{th}$ resonance as

$$p_j = \frac{\Delta A_j}{2A_j} = \frac{A_j^+ - A_j^-}{A_j^+ + A_j^-} \quad (5)$$

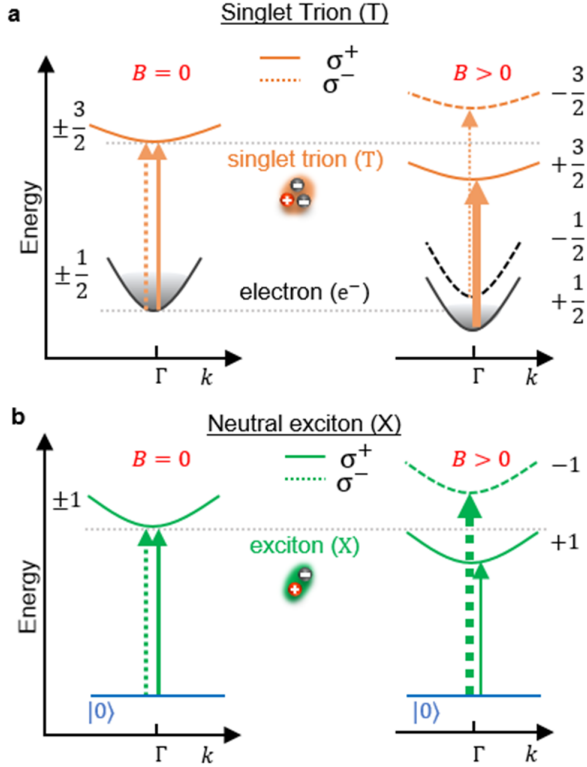

Figure 5. Schematic representation of optical transitions for (a) charged exciton T and (b) neutral exciton X, in the absence and presence of magnetic field. For $B=0$, doubly degenerate ($\sigma^{\pm}$) T and X transitions take place as excitations of the electronic states $\pm 1/2$ and vacuum state $|0\rangle$, respectively. For $B>0$, the $e^-$, X and T bands undergo a Zeeman splitting. The resulting $\sigma^+/\sigma^-$ circularly polarized transitions are depicted as solid/dashed arrows, respectively. Thickness of the arrows marks the relative optical strength of the transitions.

where $A_j = (A_j^+ + A_j^-)/2$ is the average oscillator strength. We plot $p$ in Figs. 4a and 4c for the 1s and 2s states, respectively, and notice a steady rise in magnitude for all resonances. The relative signs of $p$ for the neutral and charged excitons is one of the most important piece of evidence used to assign $T_{hh}^{2s}$ in this work. We explain this in the following.

Figure 4a suggests that for the case of 1s states i.e. $T_{hh}^{1s}$ and $X_{hh}^{1s}$ in our MOKE spectroscopy data, $p$ is of positive and negative sign, respectively. Similar trends are observed for the 2s states i.e. $T_{hh}^{2s}$ and $X_{hh}^{2s}$ (Figure 4c). Also, the polarization magnitude increases with rising magnetic field. This phenomenon is similar to the 'oscillator-strength transfer' within the exciton-trion pair as known in literature [14,28,38]. It is a fingerprint of the existence of a correlation with the electron Fermi sea mediated by the trion, i.e., the formation of the Fermi polaron, also known as Suris tetron [29]. For the exciton-electron scattering states ($X_{hh}^{1s} - e^-$), $p$ is positive as well. These observations are explained on the basis of our theoretical calculations (Figs. 4b and 4d) using the model developed in Ref. [39] as follows. A doping concentration of $0.5 \times 10^{10} \text{cm}^{-2}$ when used in our theoretical model reproduces our experimental results quite well.

To begin with, we draw a schematic showing optical transitions for trion T and neutral exciton X in the absence and presence of a magnetic field in Fig. 5 [10,13]. The initial state of a trion is the conduction band electronic state (spin $\pm 1/2$), while that of exciton, it is the vacuum state $|0\rangle$ [28,40]. The final states in the two cases are singlet trion T (spin $\pm 3/2$) and exciton X (spin $\pm 1$), respectively. For $B=0$, $\sigma^{\pm}$ transitions in the two cases are doubly degenerate. In the presence of a magnetic field, the degeneracy of the $e^-$, T and X states is lifted. This results in the Zeeman splitting of the three states. In our experiment, we observe an effective g factor with a negative sign for all cases (Figs. 3c and 3g). Figure 5 has been drawn taking this observation into account.

The trion (attractive Fermi polaron) oscillator strength in the $\sigma^{\pm}$ polarization is proportional to the density of electrons with the spin components $\pm\frac{1}{2}$ which should be picked out from the Fermi sea to form a bound singlet trion [41]. Thus, polarization of electrons in magnetic field ($B>0, g_e<0$) results in more spin-up electrons than spin-down electrons giving rise to a higher oscillator strength of $T_{hh}^{1s}$ and $T_{hh}^{2s}$ in the $\sigma^+$ polarization. This is represented qualitatively in Fig. 5a using a thicker arrow for $\sigma^+$ trion transition. At the same time, the oscillator strength of the corresponding exciton (repulsive Fermi polaron) is reduced (Fig. 5b) due to conservation of the total oscillator strength enforced by the quantum mechanical sum-rule [41,42]. Thus, the trion and exciton magneto-induced polarizations are opposite. It is also confirmed by our experimental data and our calculations presented in Fig. 4 for both 1s and 2s X − T states. Therefore, the weak shoulder observed at low energy of 2s heavy-hole exciton in our MOKE spectra belongs to the corresponding 2s trion. The sign of polarization of the trion scattering states $X_{hh}^{1s} - e^-$ in our experimental data is obviously consistent with that of bound trions and is in qualitative agreement with the theoretical calculations. It is noteworthy that all of our measurements are performed at $T = 11K$. Although, measurements at a lower temperature could result in larger trion oscillator strength and narrower spectral line widths [14,15,38], we are still able to perform data analysis with high degree of confidence and draw definite conclusions.

In summary, we observe an excited 2s trion state associated with the 2s heavy-hole exciton in a 4.2 nm wide GaAs single quantum well using MOKE spectroscopy. Our work opens many opportunities for investigations of exotic many-body systems in 2D quantum wells.

We acknowledge the financial support from the NM-ICPS of the DST, Government of India, through the I-HUB Quantum Technology Foundation (Pune, India), Project No. CRG/2022/007008 of SERB (Government of India), MoE-STARS Project No. MoE-STARS/STARS-2/2023-0912 (Government of India), and CEFIPRA CSRP Project No. 7104-2. Access to experimental facility at the Tata Institute of Fundamental Research (Mumbai, India) in the Sandip Ghosh group are deeply acknowledged. Theoretical analysis by M.M.G. was supported by the RSF project #23-12-00142. The authors also thank S. Shastri for the quantum well sample, and

# Supplemental Material

## Excited-State Trions in a Quantum Well

Sourabh Jain[1], Mikhail Glazov[2], and Ashish Arora[1,*]

[1]*Department of Physics, Indian Institute of Science Education and Research, Dr. Homi Bhabha Road, 411008 Pune, India*
[2]*Ioffe Institute, 26 Polytechnicheskaya, 194021 St. Petersburg, Russia*
*\*Email: ashish.arora@iiserpune.ac.in*

**Experimental method.** In this work, MOKE spectroscopy measurements are performed on a 4.2 nm wide GaAs single quantum well with $Al_{0.3}Ga_{0.7}As$ barriers, grown using metal-oxide vapor phase epitaxy on GaAs (001) substrate. The experimental setup (schematic in Figure S1) is based on polarization modulation spectroscopy [1,2] and is briefly described as follows. A Xenon arc lamp focuses light on the 250 μm wide entrance slit of a 0.5 m focal length monochromator. The light is dispersed from a grating (line density of 1200 lines/mm, blazed at 750 nm) which is focused on the exit slit of the monochromator. This arrangement provides a monochromatic source of 0.5 nm bandpass at 760 nm wavelength (~exciton transition wavelength). This defines the spectral resolution of the experiment. Collimated light passes through a Glan-Taylor polarizer oriented at $-45°$ to the plane of incidence. The polarized light then passes through a photoelastic modulator (PEM) with its modulation axis lying in the plane of incidence i.e. a 0° orientation. The PEM modulates the state of polarization of light at its resonance frequency $f = 42$ kHz. The PEM is located at a distance $> 1$ m above the top surface of a superconducting magnet which produces a maximum magnetic field of $B = 7$ T at the centre of a 44 mm wide bore. This keeps the PEM away from the 50 Gauss line. Light enters the bore at an angle of incidence of about 5° (near normal incidence). A cold finger of a separate continuous flow Helium cryostat resides around the center of the magnet bore, inserted from the bottom of the magnet. The quantum well sample is mounted on the top surface of the cold finger under vacuum for thermal isolation separated by an optical window. The light is focused (spot diameter ~1 mm) on the sample (size 5 mm × 5 mm) using a plano-convex lens L1 (focal length 100 mm) through a $CaF_2$ optical window. Therefore, our measurements are performed in the polar Kerr geometry. The reflected light from the sample is collimated by lens L1 (focal length 100 mm) before passing through the analyzer (Glan-Taylor polarizer situated away from the 50 Gauss line) with its pass axis oriented at 90° to the plane of incidence. Light is then detected using a photomultiplier tube connected to a lock-in amplifier and a DC voltmeter. An integration time of 50 seconds is used at every wavelength for achieving the desired signal-to-noise ratio. The reference spectrum to calculate the reflectance of the quantum well is measured on a protected silver mirror, whose reflectance in this region of wavelength is about 98.4 %.

Our present setup overcomes many challenges which are faced in performing MOKE spectroscopy in large magnetic fields:

1) Normally, optical fiber-based setups are used for magneto-optics under large magnetic fields [3,4]. However, the state of polarization of light changes after passing through an optical fiber. Therefore, we cannot perform our measurements unless all of the optical elements are mounted in free space. The present setup addresses this problem.

2) Normally, for MOKE spectroscopy, one has to avoid all possible reflections between the first polarizer and the final analyser, except the reflection from the sample itself. This is because any extra reflection disturbs the state of polarization making the MOKE spectroscopy impossible [5]. This is challenging in free-space optics setups. However, the vertical geometry of our setup providing a near normal angle of incidence solves this problem.

3) In most high-field setups, low temperature compatible optics (polarizers, wave plates, lenses etc.) is used which remains at 4 K [3,4]. However, it is not possible to perform high-precision MOKE spectroscopy when optics operates at low temperature. It is because, very slight changes in the temperature alters the birefringence of the optical elements, making the measurements unreliable. In our setup, the optics stays at room temperature since the magnetic field is produced in an open bore.

4) All polarization optics has been kept outside the 50G line which is about 1 m high from the surface of the magnet. Therefore, operations of PEM, polarizer and analyser are largely unperturbed due to stray magnetic fields.

Despite all of these improvements, our setup needs further modifications to perform MOKE spectroscopy with micron

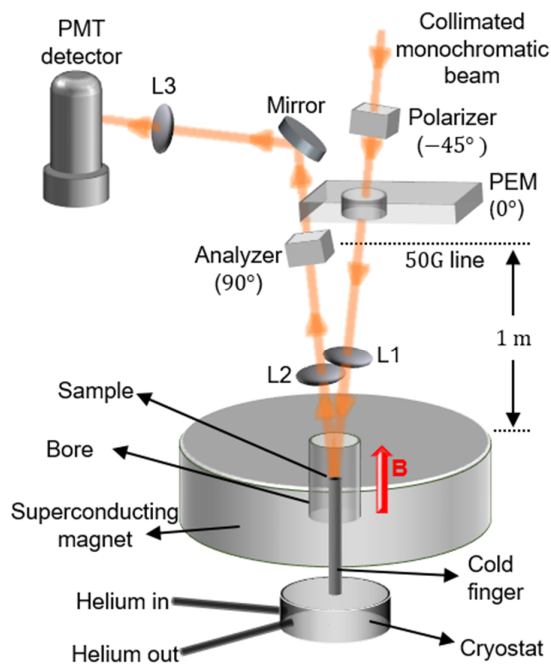

**Figure S1.** Schematic of the experimental setup.

## MOKE spectral line shapes

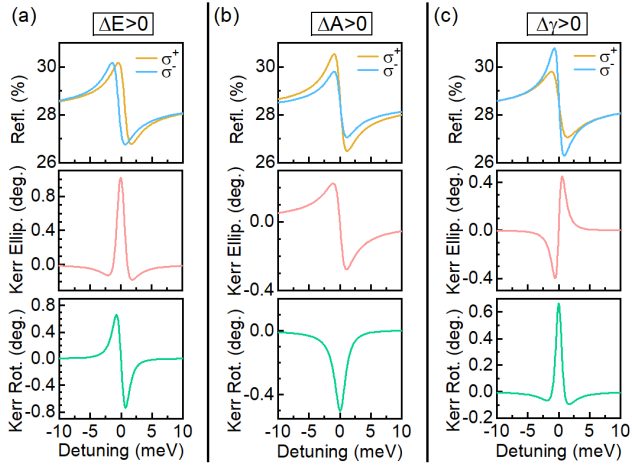

**Figure S2.** Simulated MOKE spectral line shapes assuming a dispersive reflectance line, when $\sigma^\pm$ components of the reflected light differ either in (a) transition energy $E$, i.e. non-zero Zeeman splitting $\Delta E > 0$ with $E^+ > E^-$, (b) oscillator strength $A$ i.e. polarization $p > 0$ with $A^+ > A^-$, (c) line widths i.e. $\Delta\gamma > 0$ with $\gamma^+ > \gamma^-$. Only $\Delta E \neq 0$ and $p \neq 0$ are required in the present work to fit the MOKE line shapes.

level spatial resolutions. It is because such a setup would require the placement of an objective lens just above the sample, inside large magnetic field. This will produce a Faraday rotation background of many degrees on its own. Furthermore, it is not possible to shine light at near normal incidence when using an objective lens. Therefore, such a setup will have a beamsplitter which will disturb the state of polarization of light. It becomes a part of our future work.